\documentclass{article}
\usepackage{cite}
\usepackage{graphicx}
\usepackage{dcolumn}

\begin{document}
\date{}
\title{The Hellmann-Feynman theorem and the spectrum of some Hamiltonian operators}
\author{Paolo Amore\thanks{%
e--mail: paolo@ucol.mx} \\
Facultad de Ciencias, CUICBAS, Universidad de Colima,\\
Bernal D\'{\i}az del Castillo 340, Colima, Colima, M\'exico \\
and \\
Francisco M. Fern\'andez\thanks{%
e--mail: fernande@quimica.unlp.edu.ar} \\
INIFTA, Divisi\'{o}n Qu\'{\i}mica Te\'{o}rica,\\
Blvd. 113 y 64 (S/N), Sucursal 4, Casilla de Correo 16,\\
1900 La Plata, Argentina} \maketitle

\begin{abstract}
In this short note we resort to the well known Hellmann-Feynman
theorem to prove that some non-relativistic Hamiltonian operators
support an infinite number of bound states.
\end{abstract}

\section{Introduction}

\label{sec:intro}

There has recently been some controversy about the spectrum of a rather
particular screened Coulomb potential\cite{SH21,MANY23} that was elucidated
in a later paper\cite{SCP24}. The main argument based on the
Hellmann-Feynman theorem (HFT)\cite{G32,F39} had been put forward in an
unpublished paper\cite{F23}. The purpose of this short note is the extension
of the approach just mentioned\cite{SCP24,F23} to more general cases.

In section~\ref{sec:H_general} we apply the argument based on the HFT to a
general model; in section~\ref{sec:examples} we discuss two illustrative
examples and in section~\ref{sec:conclusions} we summarize the main results
and draw conclusions.

\section{General model}

\label{sec:H_general}

The starting point of our analysis is the dimensionless Hamiltonian operator
\begin{equation}
H(\beta )=-\frac{1}{2}\nabla ^{2}-\frac{f\left( \beta /r\right) }{r},
\label{eq:H(beta)_gen}
\end{equation}
where $f(z)>0$ and $f(0)$ is finite. Under such condition it is clear that $%
H(0)$ has an infinite number of bound-state energies $E_{k}(0)<0$, $%
k=1,2,\ldots $. The transformation $(x,y,z)\rightarrow (\beta x,\beta
y,\beta z)$ leads to\cite{F20}
\begin{equation}
\beta ^{2}H(\beta )=-\frac{1}{2}\nabla ^{2}-\frac{\beta f\left( 1/r\right) }{%
r},  \label{eq:beta^2H(beta)_gen}
\end{equation}
and it follows from the HFT that
\begin{equation}
\frac{\partial }{\partial \beta }\beta ^{2}E(\beta )=-\left\langle \frac{%
f\left( 1/r\right) }{r}\right\rangle <0,  \label{eq:HFT}
\end{equation}
where $E(\beta )$ is an eigenvalue of $H(\beta )$.

Since $E_{k}(0)<0$ it stands to reason that there is a sufficiently small
value of $\beta $ such that $E_{k}(\beta )<0$ and, consequently, $\beta
^{2}E_{k}(\beta )<0$. According to the HFT (\ref{eq:HFT}) $\beta ^{2}E(\beta
)$ decreases with $\beta $ and we conclude that $E_{k}(\beta )<0$ for all $%
\beta \geq 0$. In the next section we consider two illustrative examples.

\section{Examples}

\label{sec:examples}

In what follows we apply the results of the preceding section to two
examples: the truncated Coulomb potential~\ref{subsec:TC} and the screened
Coulomb potential\ref{subsec:SC}.

\subsection{Truncated Coulomb potential}

\label{subsec:TC}

We first consider the Hamiltonian operator for the truncated Coulomb
potential\cite{MP78,LM81,P81,M81, LLALM82, DMV85, SVD85, RM89, DV90, F91,
D94, F23b}
\begin{equation}
H=-\frac{\hbar ^{2}}{2m}\nabla ^{2}-\frac{K}{\left( r^{p}+r_{0}^{p}\right)
^{1/p}},  \label{eq:H_TC}
\end{equation}
where $p>0$, $m$ is a reduced or effective mass, $r>0$ is the radial
variable and $K>0$, $r_{0}>0$ are model parameters with suitable units. If
we choose the unit of length $L=\hbar ^{2}/(mK)$ and the unit of energy $%
\epsilon =\hbar ^{2}/\left( mL^{2}\right) =mK^{2}/\hbar ^{2}$ then we obtain
the dimensionless Hamiltonian operator\cite{F20}
\begin{equation}
H=-\frac{1}{2}\nabla ^{2}-\frac{1}{\left( r^{p}+\beta ^{p}\right) ^{1/p}}=-%
\frac{1}{2}\nabla ^{2}-\frac{1}{r\left[ 1+\left( \frac{\beta }{r}\right)
^{p}\right] ^{1/p}},  \label{eq:TC_H_dim}
\end{equation}
where $\beta =r_{0}/L$ is the only relevant dimensionless parameter of the
model. Note that this Hamiltonian operator is a particular case of (\ref
{eq:H(beta)_gen}) and, consequently, it supports an infinite number of
bound-state energies.

\subsection{Screened Coulomb potential}

\label{subsec:SC}

The second example is given by Hamiltonian operator with a screened Coulomb
potential\cite{SH21, MANY23, SCP24, F23}
\begin{equation}
H=-\frac{\hbar ^{2}}{2m}\nabla ^{2}-\frac{Ae^{-B/r}}{r},  \label{eq:SC_H}
\end{equation}
where $A>0$ and $B>0$ are model parameters. In this case we choose the unit
of length $L=\hbar ^{2}/(mA)$ and the unit of energy $\epsilon =\hbar
^{2}/\left( mL^{2}\right) =mA^{2}/\hbar ^{2}$ and derive the dimensionless
Hamiltonian\cite{F20}
\begin{equation}
H=-\frac{1}{2}\nabla ^{2}-\frac{e^{-\beta /r}}{r},  \label{eq:SC_H_dim}
\end{equation}
where $\beta =B/L$ is the only relevant dimensionless parameter of the
model. Since this Hamiltonian operator is a particular case of (\ref
{eq:H(beta)_gen}) we conclude that it supports an infinite number of bound
states as argued in recent papers\cite{SCP24,F23}.

\section{Conclusions}

\label{sec:conclusions}

In this note we have shown that the HFT is extremely useful to prove the
existence of an infinite number of bound states in some quantum-mechanical
models. The main argument put forward in section~\ref{sec:H_general}
generalizes the one in earlier papers about the screened Coulomb potential%
\cite{SCP24,F23} and here we applied it also to the case of the truncated
Coulomb potential\cite{MP78,LM81,P81,M81, LLALM82, DMV85, SVD85, RM89, DV90,
F91, D94, F23b}.

\section*{Acknowledgements}

The research of P.A. was supported by Sistema Nacional de
Investigadores (M\'exico).

\end{document}